\documentclass{ws-p9-75x6-50}

\newcommand{\beq}{\begin{equation}}
\newcommand{\eeq}{\end{equation}}
\newcommand{\mc}{\multicolumn}

\begin{document}
\title{DECAY RATES OF MEDIUM-HEAVY 
${\bf \Lambda}$-HYPERNUCLEI WITHIN THE PROPAGATOR METHOD}
\author{W. M. ALBERICO, A. DE PACE, \underline{G. GARBARINO}}
\address{Dipartimento di Fisica Teorica, Universit\'a di Torino \\
and INFN, Sezione di Torino, 10125 Torino, ITALY}
\author{A. RAMOS}
\address{Departament d'Estructura i Constituents de la Mat\`eria, 
Universitat de Barcelona,\\ 08028 Barcelona, SPAIN}

\maketitle\abstracts{The $\Lambda$ decay rates in nuclei has been calculated 
in ref.~\cite{Al99} 
using the Propagator Method in Local Density Approximation.
We have studied the dependence of the widths (including the one for the 
two-body induced process $\Lambda NN\rightarrow NNN$)
on the $NN$ and ${\Lambda}N$ short range
correlations. Using a reasonable parametrization of
these correlations, as well as realistic nuclear densities and $\Lambda$ wave
functions, we reproduce, 
for the first time, the experimental non-mesonic widths 
from medium to heavy hypernuclei.}

\section{Introduction}
The study of hypernuclear physics may help in understanding some
present problems related, for instance, to aspects of weak interactions
in nuclei, or to the origin of the spin-orbit interaction in nuclei. Besides, 
it is a good instrument to study the role of quark degrees of freedom in the 
hadron-hadron interactions at short distances and
the renormalization properties of pions in the nuclear medium. 

The weak decay of hypernuclei occurs via two channels; the so called 
mesonic channel:
\beq
\Lambda \rightarrow \pi N \hspace{0.1in} (\Gamma_M) ,
\eeq
and the non-mesonic one, in which the pion emitted from the weak hadronic vertex is
absorbed by one or more nucleons in the medium, for example
\beq
\Lambda N \rightarrow NN  \hspace{0.1in} (\Gamma_1) , 
\eeq
\beq
\Lambda NN  \rightarrow  NNN \hspace{0.1in} (\Gamma_2) .
\eeq
Obviously, the non-mesonic
processes ($\Gamma_{NM}=\Gamma_1+\Gamma_2+...$)
can also be mediated by the exchange of more massive mesons. 
The non-mesonic decay is only possible in nuclei and, nowadays, the study of the
hypernuclear decay is the only practical way to get information on the
weak process
${\Lambda}N \rightarrow NN$, especially on its parity conserving part.

The free ${\Lambda}$ decay is
compatible with the ${\Delta}I=1/2$ isospin rule, which is also valid 
in other non-leptonic strangeness changing processes. 
This rule is based on the experimental observation that 
${\Gamma}_{\Lambda \rightarrow \pi^- p}/{\Gamma}_{\Lambda\rightarrow \pi^0 n}
\simeq 2$,
but it is not yet understood on theoretical grounds. 
From theoretical calculations like the one in ref.~\cite{Os93}
and from experimental measurements \cite{Os98} there is some evidence that the
${\Delta}I=1/2$ rule is broken in the nuclear mesonic decay. 
However, this is essentially due to shell
effects and might not be directly related to the weak process. 
A recent estimate of $\Delta I=3/2$ contributions to the
$\Lambda N \to N N$ reaction \cite{Pa98} found  
moderate effects on the hypernuclear decay rates.
In the present calculation of the decay rates in nuclei we will assume this rule as valid. 
The momentum of the final nucleon
in ${\Lambda}\rightarrow \pi N$ is about $100$ MeV for ${\Lambda}$ at rest,
so this process is suppressed by the Pauli principle, particularly in heavy nuclei. 
It is strictly forbidden in infinite nuclear matter 
(where $k_F^0\simeq$ 270 MeV), but in finite nuclei it can 
occur because of three important effects: 1) in nuclei 
the Pauli blocking is less effective and the hyperon 
has a momentum distribution which allows larger momenta for the final nucleon,
2) the final pion feels an attraction by the medium such that for fixed
momentum it has an energy smaller than the free one and consequently, 
the final nucleon again has more chance to come out above 
the Fermi surface, and 3) on the nuclear surface the local Fermi 
momentum is smaller than $k_F^0$ and 
favours the decay. Nevertheless, the mesonic width decreases fast as 
the mass number $A$ of the hypernucleus increases \cite{Os93}. 
The mesonic rate is very sensitive to the pion self-energy,
so that from the study of the mesonic decays it could be possible to extract
important information on the pion-nucleus optical potential, which we do not
know, nowadays, in a complete form.

The final nucleons in the non-mesonic process
${\Lambda}N\rightarrow NN$ come out with a momentum
$\simeq 420$ MeV, so that this decay is not 
forbidden by the Pauli principle. On the contrary, apart from the $s$-shell 
hypernuclei,
it dominates over the mesonic decay. The non-mesonic channel is characterized by
large momentum transfers, so that the details of the nuclear structure do not 
have a substantial influence
while the $NN$ and ${\Lambda}N$ short range correlations turn out to be very 
important. There appears to be an
anticorrelation between mesonic and non-mesonic decay modes: indeed the total
lifetime is fairly constant from light to heavy hypernuclei~\cite{Os98,Co90}: 
${\tau}_{exp}=(0.5\div 1)\, {\tau}_{free}$.

Nowadays, the main problem concerning the weak decay rates 
is to reproduce the experimental values for the ratio ${\Gamma}_n/{\Gamma}_p$
between the neutron and the proton induced widths 
${\Lambda}n\rightarrow nn$ and ${\Lambda}p\rightarrow np$. All theoretical
calculations underestimate the experimental data for all the considered 
hypernuclei \cite{Os98,Pa98,Pa97,Ok98,Du96,It95}:
\beq
\left\{\frac{{\Gamma}_n}{{\Gamma}_p}\right\}^{Th}\ll
\left\{\frac{{\Gamma}_n}{{\Gamma}_p}\right\}^{Exp}
\hspace{0.8in}
0.5\le \left\{\frac{{\Gamma}_n}{{\Gamma}_p}\right\}^{Exp}\le 2 .
\eeq
In the One Pion Exchange (OPE) approximation the values for this 
ratio are $0.1\div 0.2$. On the other hand, the
OPE model has been able to reproduce the 1-body stimulated non-mesonic rates 
${\Gamma}_{1}={\Gamma}_n+{\Gamma}_p$ for light and medium 
hypernuclei~\cite{Pa97,Ok98,It95}. In order to solve this puzzle many attempts have
been made up to now, but without success. Among these we recall the 
inclusion in the ${\Lambda}N\rightarrow NN$
transition potential of mesons heavier than the pion 
\cite{Pa97,Du96,It95}, the inclusion of contributions that
violate the ${\Delta}I=1/2$ rule \cite{Pa98} and the description of the 
short range baryon-baryon interaction in terms of quark degrees of freedom
\cite{Ok98}. This last calculation is the only 
one which has found a fairly large (but not sufficient) increase of the neutron 
to proton ratio with respect to the OPE one.
However, this calculation is carried out
only for $s$-shell hypernuclei; moreover, the employed quark-lagrangian 
does not reproduce
the experimental ratio between the ${\Delta}I=1/2$ and ${\Delta}I=3/2$ 
transition amplitudes for the ${\Lambda}$ free decay.

The analysis of the ratio ${\Gamma}_n/{\Gamma}_p$ is influenced by the
2-nucleon induced process ${\Lambda}NN\rightarrow NNN$.
By assuming that the meson produced in the weak
vertex is mainly absorbed by a strongly correlated neutron-proton pair, 
the 3-body process turns out to be
${\Lambda}np\rightarrow nnp$, so that 
a considerable fraction of the measured neutrons 
could come from this channel and not only from the
${\Lambda}n\rightarrow nn$ and ${\Lambda}p\rightarrow np$ ones. 
In this way it might be
possible to explain the large experimental ${\Gamma}_n/{\Gamma}_p$ ratio,
which originally has been analyzed without taking into account 
the 2-body stimulated process. 
Nevertheless,
the situation is far from being clear and simple~\cite{Os98,Ra95}. 
In fact, ${\Gamma}_n/{\Gamma}_p$ is sensitive to the energy
spectra of the emitted nucleons, whose calculation also requires a careful
treatment of the final state interactions of the nucleons. 
In ref.~\cite{Ra97} the energy distributions 
were calculated using a Monte Carlo simulation to describe 
the final state interactions. 
A direct comparison of those spectra with 
the experimental ones seems to
favour large ${\Gamma}_n/{\Gamma}_p$ ratios, in strong disagreement 
with the OPE predictions. However, it was also pointed out the 
convenience of measuring the number of protons per decay event. 
In fact, this observable gives a more reliable
$\Gamma_n/\Gamma_p$ ratio and it is less sensitive 
to details of the Monte Carlo
simulation accounting for the final shape of the spectra.

\section{Decay widths}
\label{Decay widths}
We present here the results of a new evaluation \cite{Al99} of 
the decay rates for medium to heavy $\Lambda$-hypernuclei 
based on the Propagator Method introduced in ref.~\cite{Os85}. The parameters
of the model were adjusted to reproduce the non-mesonic width of 
$^{12}_{\Lambda}$C. Then the decay rates for heavier hypernuclei have
been predicted and compared, when possible, with the experiment.
\subsection{The model}
\label{themodel}

The propagator technique, estensively explained in 
refs.~\cite{Al99,Os85}, provides a unified picture
of the different decay channels and it is equivalent to the standard
Wave Function Method \cite{Os94}.
The calculation of the widths is performed in nuclear matter, and
then extended to finite nuclei via the Local Density Approximation (LDA).

The starting point for the calculation is the ${\Lambda}\rightarrow \pi N$ 
effective lagrangian, parametrized in the form:
\beq
\label{lagran}
{\cal L}_{{\Lambda}\pi N}=G m_{\pi}^2\overline{\psi}_N(A+B\gamma_5)
{\vec \tau} \cdot {\vec \phi}_{\pi}{\psi}_{\Lambda}+h.c. ,
\eeq
where the values of the weak coupling constants 
$G$, $A$, $B$ are fixed on the observables of the free ${\Lambda}$ decay. 
In order to enforce the ${\Delta}I=1/2$ rule, in eq.~(\ref{lagran}) 
the hyperon is assumed to be an isospin spurion with $I_z=-1/2$.
The decay rate of a ${\Lambda}$ in nuclear matter is obtained from the 
imaginary part of the ${\Lambda}$ self-energy through the relation:
\beq
\label{Gamma}
{\Gamma}_{\Lambda}=-2{\rm Im}{\Sigma}_{\Lambda} .
\eeq
The $\Lambda$ self-energy, ${\Sigma}_{\Lambda}$, contains an infinite series of diagrams
in which the pion emitted in the $\Lambda$ vertex couples to {\sl 1p-1h}, {\sl 2p-2h}, etc
excitations, giving rise to the different decay channels when 
they are cut in all the possible ways.
The structure of the decay width in nuclear matter is the following:
\beq
\label{Sigma2}
{\Gamma}_{\Lambda}(\vec k)\propto \int d\vec q . . .
{\theta}(\mid \vec k- \vec q \mid -k_F)
{\theta}(k_0-E_N(\vec k-\vec q)-V_N)
{\rm Im}G_{\pi}(q)\mid _{q_0=k_0-E_N(\vec k-\vec q)-V_N} ,
\eeq
where
\beq
\label{proppion}
G_{\pi}(q)=\frac{1}{q_0^2-\vec q^2-m_{\pi}^2-{\Sigma}_{\pi}^*(q)} ,
\eeq
is the pion propagators in nuclear matter, ${\Sigma}_{\pi}^*$ being its
proper self-energy; $k_F$ is the Fermi momentum, $E_N$ is the nucleon total free
energy and $V_N$ is the nucleon binding energy. 
Moreover, $k_0=E_{\Lambda}(\vec k)+V_{\Lambda}$, is the ${\Lambda}$ energy,
containing a binding term (which we take from the experiment). 

The decay widths in finite nuclei are obtained in LDA;
a local Fermi sea of nucleons, related to the nuclear density by
\beq
\label{local}
k_F(\vec r)=\left\{\frac{3}{2}{\pi}^2\rho 
(\vec r)\right\}^{1/3} ,
\eeq
is introduced. Besides, the nucleon binding potential $V_N$ also 
becomes $r$-dependent in LDA.
The decay width in finite nuclei is then obtained through the relations:
\beq
\label{local1}
{\Gamma}_{\Lambda}(\vec k)=\int d\vec r \mid {\psi}_{\Lambda}(\vec r)\mid ^2
{\Gamma} _{\Lambda}[\vec k,\rho (\vec r)] , \hspace{0.2in}
{\Gamma}_{\Lambda}=\int d\vec k \mid \tilde{\psi}_{\Lambda}(\vec
k)\mid^2{\Gamma}_{\Lambda}(\vec k) ,
\eeq
where ${\psi}_{\Lambda}$ ($\tilde{\psi}_{\Lambda}$) is the ${\Lambda}$ 
wave function in the coordinate (momentum) space.

\subsection{Results}
\label{results}

In ref.~\cite{Al99} we have studied the influence of the short range
correlations and the $\Lambda$ wave function on the decay width of
$^{12}_\Lambda$C, which we have
used as a testing ground for the theoretical framework in order to fix the
parameters of our model. 

The baryon-baryon 
correlations are governed by the Landau parameters $g^{\prime}$ and 
$g^{\prime}_{\Lambda}$.
The choice of these parameters which reproduce the experimental
non-mesonic rate of $^{12}_{\Lambda}C$ is $g^{\prime}=0.8$,
$g^{\prime}_{\Lambda}=0.4$. Neglecting the contribution of the 
2-body induced decay channel (namely in ring approximation),
with the same $g^{\prime}_{\Lambda}$, we find that $g^{\prime}=0.7$ 
should de used. This value is in agreement with the fenomenology of the
$(e,e')$ quasi-elastic scattering \cite{Os82}.

The ${\Lambda}$ wave function is obtained from a Wood-Saxon well which
exactly reproduces the first two experimental $\Lambda$ single particle eigenvalues 
($s$ and $p$ levels). The numerical results for $^{12}_{\Lambda}C$ 
are shown in table~\ref{wf sens},
where they are compared with the
experimental data from BNL \cite{Sz91} and KEK \cite{No95,Bh98,Sa98}.
\begin{table}
\caption{Decay rates for $^{12}_{\Lambda}$C} 
\vspace{0.2cm}
\begin{center}
\label{wf sens}
\begin{tabular}{|c|c c c c|c}
\hline
\mc {1}{|c|}{}&
\mc {1}{c}{Calculation} &
\mc {1}{c}{BNL \cite{Sz91}} &
\mc {1}{c}{KEK \cite{No95}} &
\mc {1}{c|}{KEK New \cite{Bh98,Sa98}} \\ \hline
${\Gamma}_M$     & 0.25 & $0.11\pm 0.27$ & $0.36\pm 0.15$ 
& $>0.11$\\
${\Gamma}_1$     & 0.82 &          &          &    \\
${\Gamma}_2$     & 0.16 &          &          &    \\
${\Gamma}_{NM}$  & 0.98 &  $1.14\pm 0.20$ & $0.89\pm 0.18$
& $<1.03$ \\
${\Gamma}_T$ & 1.23 &  $1.25\pm 0.18$ & $1.25\pm 0.18$
& $1.14\pm 0.08$ \\ 
\hline 
\end{tabular}
\end{center}
\end{table}
In addition
to the Wood-Saxon wave function that reproduces the $s$ and $p$ levels, we 
also made a test using the harmonic oscillator wave
function with an ''experimental'' frequency $\omega$
obtained from the {\sl s-p} experimental energy shift, the 
Wood-Saxon wave function of Dover {\sl et al.}\ \cite{Do88} and
the microscopic wave function calculated from a non-local self-energy using
a realistic $\Lambda N$ interaction in ref.~\cite{Po98}.
We have found that all these different $\Lambda$ wave functions
give rise to total decay widths which may differ at most by 15\%. 

Then, using Wood-Saxon wave functions that reproduce the $s$ and $p$ 
$\Lambda$-levels and the Landau parameters $g^{\prime}=0.8$,
$g^{\prime}_{\Lambda}=0.4$, we have extended the 
calculation to heavier hypernuclei. 
We note that, in order to reproduce 
the experimental $s$ and $p$ levels for the hyperon we must
use potentials with a nearly 
constant depth, around $28\div 32$ MeV, from medium to heavy hypernuclei.

\begin{figure}[p]
\epsfysize=13cm
\epsfbox{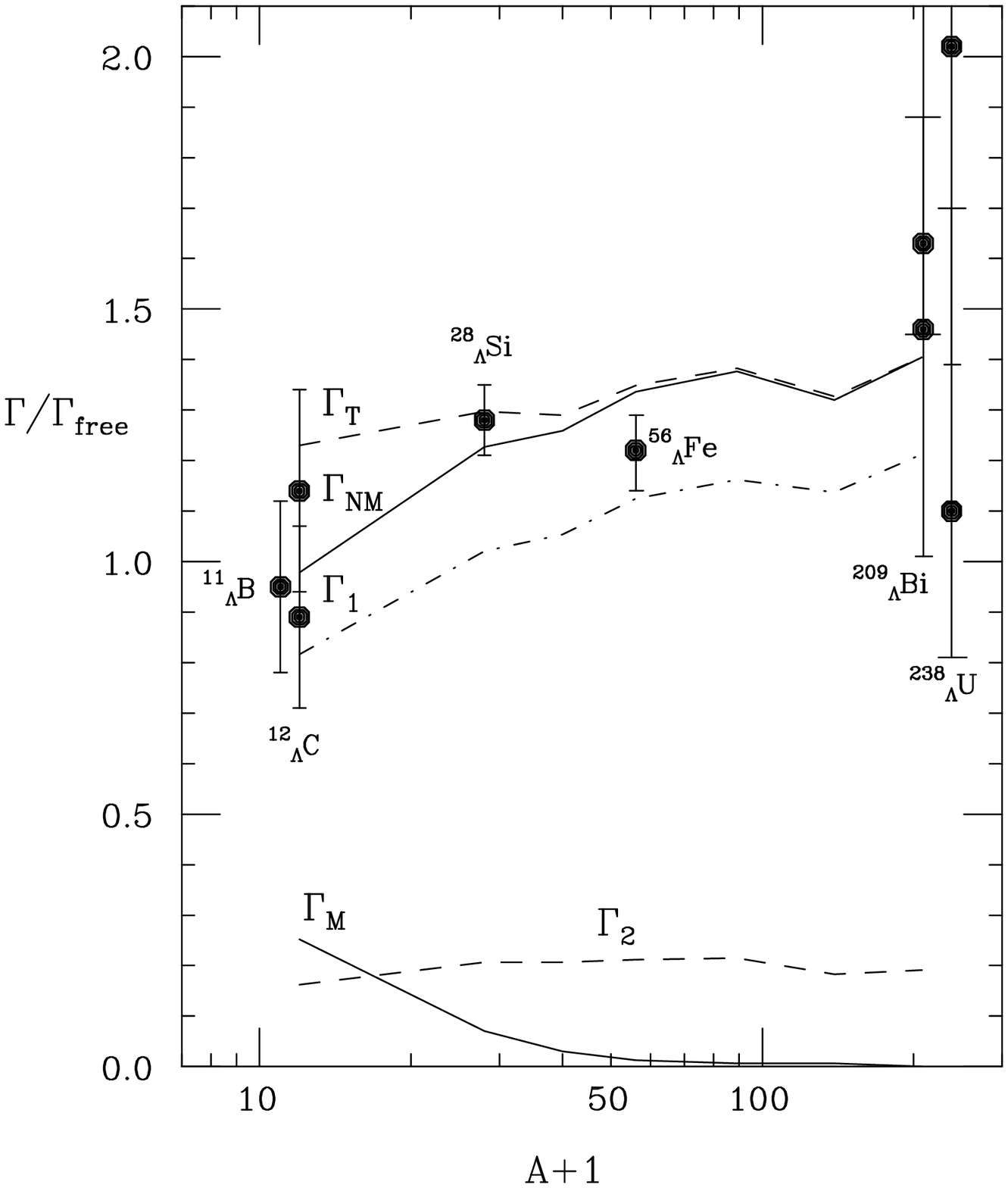}
\caption{${\Lambda}$ decay widths in finite nuclei as a function of
the mass number $A$.  
\label{satu}}
\end{figure}

Our results are shown in table~\ref{sat}. 
\begin{table}
\caption{Decay rates of medium-heavy $\Lambda$-hypernuclei}
\vspace{0.2cm}
\begin{center}
\label{sat}
\begin{tabular}{|c|c c c c|}
\hline
\mc {1}{|c|}{$ ^{A+1}_{\Lambda}Z$} &
\mc {1}{c}{${\Gamma}_M$} &
\mc {1}{c}{${\Gamma}_1$} &
\mc {1}{c}{${\Gamma}_2$} &
\mc {1}{c|}{${\Gamma}_{T}$} \\ \hline
$ ^{12}_{\Lambda}$C    & 0.25             & 0.82 & 0.16 & 1.23 \\
$ ^{28}_{\Lambda}$Si   & 0.07             & 1.02 & 0.21 & 1.30 \\
$ ^{40}_{\Lambda}$Ca   & 0.03             & 1.05 & 0.21 & 1.29 \\
$ ^{56}_{\Lambda}$Fe   & 0.01             & 1.12 & 0.21 & 1.35 \\
$ ^{89}_{\Lambda}$Y    & $6\cdot 10^{-3}$ & 1.16 & 0.22 & 1.38 \\
$ ^{139}_{\Lambda}$La  & $6\cdot 10^{-3}$ & 1.14 & 0.18 & 1.33 \\
$ ^{208}_{\Lambda}$Pb  & $1\cdot 10^{-4}$ & 1.21 & 0.19 & 1.40 \\
\hline
\end{tabular}
\end{center}
\end{table}
The mesonic rate rapidly vanishes by increasing the mass number 
$A$. This is well known and it is related to the decreasing phase space 
allowed for the mesonic channel, and to smaller overlaps between 
the ${\Lambda}$ wave function ${\psi}_{\Lambda}$ and the nuclear surface,
as $A$ increases.
The 2-body induced decay is rather independent of the hypernuclear dimension
and it is about 15\% of the total width. The total width is also
nearly constant with $A$, as we already know from the experiment.
In fig.~\ref{satu} we compare the results from table~\ref{sat}
with recent (after 1990) experimental data for non-mesonic decay
\cite{Sz91,No95,Bh98,Ku98,Ar93,Oh98}. 
We remind that the data for nuclei heavier than 
$^{12}_{\Lambda}$C refer to the {\it total} 
width. The theoretical results are in good agreement with the data over the 
whole hypernuclear mass range explored. Moreover, we also 
see that the saturation of the ${\Lambda}N\rightarrow NN$ interaction 
in nuclei appears to be well reproduced.

The main open problem in the study of weak hypernuclear decays is to
understand the large experimental value of the ratio $\Gamma_n/\Gamma_p$.
However, we have to remind that the data for $\Gamma_n/\Gamma_p$ have a large
uncertainty and they have been analyzed without taking
into account the 3-body decay mechanism. 
The study of ref.~\cite{Ra95} showed that, even if the three body
reaction is only about 15 \% of the total decay rate, this mechanism 
influences the analysis of the data determining the ratio $\Gamma_n/\Gamma_p$.
Energy spectra of neutrons and protons from the non-mesonic decay
mechanisms were calculated in ref. \cite{Ra97}. The momentum distributions of
the primary nucleons were determined from the Propagator Method 
and a subsequent Monte Carlo simulation was used to account for the final state
interactions. It was shown that the shape of the proton spectrum is sensitive 
to the ratio $\Gamma_n/\Gamma_p$. In fact, the protons from the three-nucleon
mechanism appeared mainly at low energies, while those from the
two-body process peaked around 75 MeV. Since the experimental spectra show a
fair
amount of protons in the low energy region they would favour a relatively large
three-body decay rate or, conversely, a reduced number of protons from the
two-body process. Consequently, the experimental spectra were compatible with
values for $\Gamma_n/\Gamma_p$ around $2 \div 3$, in strong contradiction with
the present theories.
 
Within the Propagator Method with modified parameters and realistic 
$\Lambda$ wave functions
we have then generated new nucleon spectra for various hypernuclei
using the Monte Carlo simulation of ref.~\cite{Ra97}.
Although the new non-mesonic widths are sizably reduced
(by about 35\%) with respect to those of refs.~\cite{Ra95,Ra97}, 
the resulting nucleon spectra are practically identical. 
The reason is that the ratio $\Gamma_2/\Gamma_1$ is
essentially the same in both models, and the 
momentum distributions for the primary emitted nucleons are also very similar.
As a consequence, the conclusions drawn in ref.~\cite{Ra97} still hold.

Therefore, the
origin of the discrepancy between theory and experiment for the ratio
$\Gamma_n/\Gamma_p$ still
needs to be resolved. However, we must notice that on the experimental side,
although new spectra are now available \cite{No95,Bh98}, they have
not been corrected for energy losses inside the target and detector, so a
direct comparison with the theoretical predictions is not yet possible.
Attempts to incorporate these corrections by combining a theoretical model 
for the nucleon rescattering in the nucleus with a simulation of the energy 
losses in the experimental set-up are now being pursued \cite{Ou99}.
On the other hand, a forward step towards a clean extraction of the ratio
$\Gamma_n/\Gamma_p$ would be obtained if
the nucleons which come out from the different
non-mesonic processes, $\Lambda N \to NN $ and $\Lambda NN \to NNN$ were
disentangled \cite{Ze98}.

\section*{Acknowledgments}We would like to thank H. Noumi and H. Outa for 
discussions and for giving us detailed information about the experiments.


\end{document}